\documentclass[english,twocolumn,showpacs,preprintnumbers,prl]{revtex4}
\usepackage[T1]{fontenc}
\usepackage[latin1]{inputenc}
\usepackage{verbatim}
\usepackage{graphicx}
\usepackage{amssymb}

\makeatletter

\providecommand{\boldsymbol}[1]{\mbox{\boldmath $#1$}}

\usepackage{graphics}

\usepackage{psfig}
\usepackage{color}

\usepackage{babel}
\makeatother
\begin{document}

\title{Ensemble inequivalence, bicritical points and azeotropy for generalized
Fofonoff flows}

\author{Antoine Venaille and Freddy Bouchet}

\thanks{Freddy.Bouchet@inln.cnrs.fr}

\affiliation{Laboratoire des Écoulements Géophysiques et Industriels, UJF, INPG,
CNRS ; BP 53, 38041 Grenoble, France and Institut Non Linéaire de
Nice , CNRS, UNSA, 1361 route des lucioles, 06 560 Valbonne - Sophia
Antipolis, France}

\date{{\normalsize \today}}

\begin{abstract}
We present a theoretical description for the equilibrium states of
a large class of models of two-dimensional and geophysical flows,
in arbitrary domains. We account for the existence of ensemble inequivalence
and negative specific heat in those models, for the first time using
explicit computations. We give exact theoretical computation of a
criteria to determine phase transition location and type. Strikingly,
this criteria does not depend on the model, but only on the domain
geometry. We report the first example of bicritical points and second
order azeotropy in the context of systems with long range interactions. 
\end{abstract}


\maketitle
In many fields of physics, the particle or fields dynamics is not
governed by local interactions. For instance for self gravitating
stars in astrophysics \cite{Chavanis:2002_PhysRevE_Phasetransition_SGS,IspolatovCohen:2001_PRE_PhaseTransition_SGS},
for vortices in two dimensional and geophysical flows \cite{Miller:1990_PRL_Meca_Stat,Robert:1991_JSP_Meca_Stat,SommeriaRobert:1991_JFM_meca_Stat},
for unscreened plasma or models describing interactions between waves
and particles \cite{MarcorDoveilElskens:2005_PRL_Wave_Particle},
the interaction potential is not integrable \cite{DRAW:2002_Houches}.
Recently, a new light was shed on the equilibrium statistical mechanics
of such systems with long range interactions : there has been a mathematical
characterization of ensemble inequivalence \cite{EllisHavenTurkington:2000_Inequivalence},
a study of several simple models \cite{BarreMukamelRuffo:2001_PRL_BEC,CosteniucEllisTouchette:2005},
and a full classification of phase transitions and of ensemble inequivalence
\cite{BB:2005_JSP} %
\begin{comment}
and the appearance of a very useful technique, the large deviation
theory \cite{BBDR:2005_JSP}
\end{comment}
{}. %
\begin{comment}
Another motivation comes from the the understanding that the broad
spectrum of applications should be considered simultaneously since
significant advances were performed independently in the different
domains.
\end{comment}
{}%
\begin{comment}
Currently the dynamics and the out of equilibrium study \cite{BaldovinOrlandini:2006_PRL_ThermalBath,AntoniazziEFR:2006_EPJB_Meca_Stat_Vlasov}
or the study of short range interactions on such systems \cite{CampaGMR:2006_PhysicaA_ShortandLongRange}
are also extremely active fields.
\end{comment}
{}

\begin{figure}
\resizebox{8truecm}{!}{\includegraphics{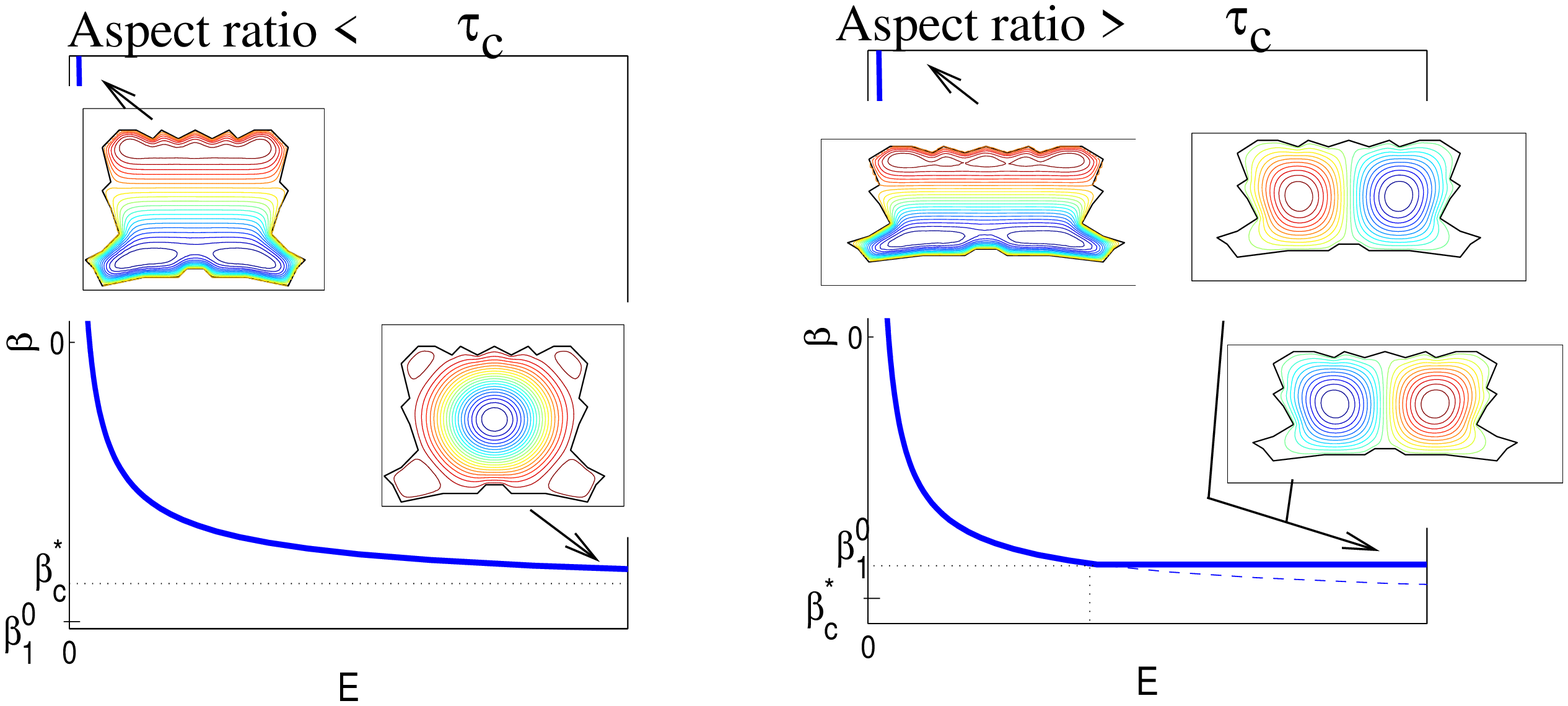}}

\caption{Geometry governed phase transition : second order phase transition
from Fofonoff modes in a domain with axial symmetry to dipole solutions,
breaking the symmetry, only when the domain is sufficiently stretched
$\left(\tau>\tau_{C}\right).$}

\label{fig:Phase_Transition_General_domain} 
\end{figure}

One of the promising field of application for the statistical mechanics
of systems with long range interactions, is the statistical prediction
of large scale geophysical flows. For instance, the structure of Jupiter's
troposphere has been successfully explained using the Robert-Sommeria-Miller
(RSM) equilibrium theory \cite{BS:2002_JFM} \cite{TurkingtonMHD:2001_PNAS_GRS}%
\begin{comment}
the Great Red Spot has been precisely and quantitatively modeled \cite{BS:2002_JFM}
and its position in the south hemisphere explained \cite{TurkingtonMHD:2001_PNAS_GRS}
\end{comment}
{}. One of the major scope of this field is to go towards earth ocean
applications. All textbook in oceanography present the Fofonoff flows
which have played an important historical role in that field %
\begin{comment}
\cite{Pedlosky:1998_OceanCirculationTheory}
\end{comment}
{}\cite{Fofonoff:1954_steady_flow_frictionless}. In this letter, we
propose a theoretical description of such flows in the context of
the statistical theories which, for the first time, relates its properties
to phase transitions (see Fig. \ref{fig:Phase_Transition_General_domain}),
negative specific heat and ensemble inequivalence. %
\begin{comment}
We use the expression generalized Fofonoff flows because, by contrast
with the original study of Fofonoff \cite{Fofonoff:1954_steady_flow_frictionless},
we consider a variational problem and we use a much larger class of
models.
\end{comment}
{}

One of the striking features of the equilibrium theory of systems
with long range interactions is the generic existence of negative
specific heat. %
\begin{comment}
: the temperature decreases when the energy increases. 
\end{comment}
{}This strange phenomena is possible as a consequence of the lack of
additivity of the energy and is related to the inequivalence between
the microcanonical and canonical ensemble of statistical physics.
This was first predicted in the context of astrophysics \cite{LyndenBell:1968_MNRAS}.
For two dimensional flows, existence of such inequivalence has been
matematically proven for point vortices \cite{CagliotiLMP:1995_CMP_II(Inequivalence)}
(without explicit computation)\textbf{,} and numerically %
\begin{comment}
mathematically proved for the point vortex model \cite{CagliotiLMP:1995_CMP_II(Inequivalence)}
\end{comment}
{}observed in a particular situation of a Quasi-Geostrophic (QG) model
\cite{EllisHavenTurkington:2002_Nonlinearity_Stability}, and in a
Monte Carlo study of points vortices in a disk \cite{Smith_ONeil:1990_phys_fluid}.
One of the novelty and achievement of the current work, is that we
predict such ensemble inequivalence, with exact theoretical computation
of the associated phase transitions, for a very large class of models
including the Euler equation or  QG models%
\begin{comment}
, whatever the topography or the Rossby radius of deformation
\end{comment}
{}. %
\begin{comment}
Besides the observation of ensemble inequivalences, the theoretical
arguments allows for a clear explanation of it. 
\end{comment}
{}

In the context of systems with long range interactions, a classification
of phase transitions associated to ensemble inequivalence has been
proposed \cite{BB:2005_JSP}. Some of the transitions predicted have
never been observed, neither in models, nor in real physical systems.
%
\begin{comment}
for instance negative temperature jumps at a phase transition when
the energy is increased.
\end{comment}
{}One of the main interest of the current work, is the finding of two
examples of such unobserved phase transitions : bicritical points
(a bifurcation from a first order phase transition towards two second
order phase transitions) and second order azeotropy (the simultaneous
appearance of two second order phase transitions at a bifurcation).
We prove that those phase transitions are governed by the geometry
of the domain in which the flow takes place. %
\begin{comment}
We explain that such geometry-governed phase transitions does not
exist in systems with short range interactions.
\end{comment}
{}We explain how to easily compute the transition point for any domain
geometry. %
\begin{comment}
Strikingly the critical value for the transition does not depend on
the model.
\end{comment}
{} %
\begin{comment}
One of the present aim of the research in systems with long range
interactions is to find experimental setups in which ensemble inequivalence
and negative specific heat could be observed in laboratories. This
very interesting issue is discussed in the conclusion ; we explained
why such generalized Fofonoff flows are serious candidates to first
observe ensemble inequivalence and its stringent physical consequences.
\end{comment}
{}

\paragraph{Euler and QG equations}

This letter describes phase transitions existing in a broad ensemble
of models : 2D Euler flows, one layer QG models \textbf{}in a closed
domain $\mathcal{D}$. The common character of all of those models
comes from the fact that they can all be expressed as a quasi-2D transport
equation $\partial_{t}q+\boldsymbol{u\cdot\nabla}q=0$. For the one
layer QG model, the potential vorticity (PV) $q=\Delta\psi-\psi/R^{2}+h$
is a scalar ; $h(x,y)$ is the topography and $R$ is the Rossby radius
of deformation. The velocity field is related to $q$ via $\psi$
: $\mathbf{u}=\mathbf{e}_{z}\times\boldsymbol{\nabla}\psi$. The case
$h=0$, $R=+\infty$ corresponds to the Euler equation. %
\begin{comment}
The term $q$ can as well be a vector in the case of multi level QG
equations.
\end{comment}
{} %
\begin{comment}
with each component satisfying $q_{(i)}=\Delta\psi_{(i)}-(2\psi_{(i)}-\psi_{(i-1)}-\psi_{(i+1)})/R_{(i)}^{2}+h_{(i)}$
\end{comment}
{}

For all of these models we use an impermeability condition at the
boundaries. %
\begin{comment}
For instance we may use $\psi=0$ at the boundary in the one layer
QG model.
\end{comment}
{}The energy of such systems can always be written $\mathcal{E}[q]=-\frac{1}{2}\langle\mathcal{O}[q-h],q-h\rangle$
where $\mathcal{O}$ is a symmetric linear operator.%
\begin{comment}
and where $\left\langle f,g\right\rangle =\int_{\mathcal{D}}f(x,y)g(x,y)dxdy$.
\end{comment}
{}For instance, $\mathcal{E}=\frac{1}{2}\langle\left(\nabla\psi\right)^{2}+\psi^{2}/R^{2}\rangle$
in the 1-1/2 layer QG model, which corresponds to $\mathcal{O}=(\Delta-R^{-2})^{-1}$%
\begin{comment}
(with the choice $\psi=0$ at boundaries)
\end{comment}
{}. In all these models, both energy and circulation $\mathcal{C}[q]=\left\langle q\right\rangle $
are conserved quantities.

\paragraph*{Variational problem}

In the following we consider the solutions of the variational problem
:\begin{equation}
S(E,\Gamma)=\max_{q}\left\{ \mathcal{S}[q]\ |\ \mathcal{E}[q]=E\ \&\ \mathcal{C}[q]=\Gamma\right\} ,\label{eq:ProblemeVariationnel}\end{equation}
where $\mathcal{S}$ is the entropy of the PV field $q$ : $\mathcal{S}[q]\ =\left\langle s(q)\right\rangle $,
with $s$ a concave function, and $S\left(E,\Gamma\right)$ is the
equilibrium entropy. Such a variational problem may be interpreted
in two ways. First, in the Robert-Sommeria-Miller statistical mechanics
of the Euler equation, one obtain a much more complex variational
problem which involves the usual Maxwell-Boltzmann entropy and which
is constrained on the whole initial vorticity distribution. One can
prove that if this vorticity distribution is treated canonically,
then the statistical equilibrium verify the variational problem (\ref{eq:ProblemeVariationnel})
\cite{Bouchet:2007_condmat}. %
\begin{comment}
, where the shape of $s$ is given by the Lagrange parameters $\Pi\left(\sigma\right)$
associated to the conservation of the vorticity (see \cite{Bouchet:2007_condmat}).
\end{comment}
{}Moreover any solution to (\ref{eq:ProblemeVariationnel}) is a RSM
equilibrium \cite{Bouchet:2007_condmat}. An alternative interpretation
is to assume directly that, in some physical situations, one has to
consider the most probable state with respect to a prior vorticity
distribution%
\begin{comment}
$\Pi\left(\sigma\right)$
\end{comment}
{}, which can be related to the shape of $s$ \cite{EllisHavenTurkington:2000_Inequivalence}.%
\begin{comment}
In both cases $s$ is the Legendre-Fenchel transform of $\Pi\left(\sigma\right)$.
\end{comment}
{} %
\begin{comment}
For instance, in the case of Gaussian $\Pi\left(\sigma\right)$, one
obtains a quadratic entropy : $\mathcal{S}[q]\ =\left\langle -q^{2}/2\right\rangle $.
\end{comment}
{} 

To compute critical points of the variational problem (\ref{eq:ProblemeVariationnel}),
we introduce two Lagrange parameters $\beta$ and $\gamma$ associated
respectively with the energy and the circulation conservation. These
critical points are stationary solutions for the initial transport
equation : $q=f(\psi)$, with $f\left(\psi\right)=s'^{-1}\left(-\beta\psi+\gamma\right)$.
In all of the following, we study the case of a quadratic entropy
$\mathcal{S}[q]\ =\left\langle -q^{2}/2\right\rangle $. %
\begin{comment}
This particular case plays a very special role, as it
\end{comment}
{}This leads to a linear vorticity-stream function relationship ($f\left(\psi\right)=\beta\psi-\gamma$).
The original Fofonoff solution corresponds to the particular case
$\beta\gg1$ and $h(x,y)=y$\textbf{.} Among the states we study,
only the ones with $\beta$ lower than the phase transition value
have already been described as Fofonoff flows or in the context of
the Kraichnan statistical mechanics \cite{Kraichnan:1975_statisical_dynamics_2D_flow,SalmonHollowayHendershott:1976_JFM_stat_mech_QG}.
%
\begin{comment}
The interpretation we propose here is rather different, as all the
state we will describe are equilibriums of the RSM theory or of its
Turkington variance. Moreover, as will be seen, the thermodynamical
description of such states has not really been considered so far. 
\end{comment}
{}In the context of the RSM statistical mechanics, in the case of Euler
equation, Chavanis and Sommeria \cite{ChavanisSommeria:1996_JFM_Classification}
found a criteria for the existence of a transition from a monopole
to a dipole when increasing the energy (as in figure \ref{fig:Phase_Transition_General_domain}).
By using a different method (by solving directly the variational problem),
we generalize these results to a wide class of model, analyze for
the first time ensemble inequivalence, find unobserved phase transitions
and establish the relations of these to Fofonoff flows.

\paragraph{Dual variational problems}

Our problem is to find the minimum of a quadratic functional (\ref{eq:ProblemeVariationnel}),
taking into account the constraints on circulation (linear) and energy
(quadratic with possibly a linear contribution). This will be referred
as the microcanonical problem. Dealing with unconstrained variational
problems is much easier than dealing with constrained ones: solutions
for a variational problem are necessarily solutions for a more constrained
dual problem \cite{EllisHavenTurkington:2000_Inequivalence}. Moreover,
when all the possible constraint values are achieved in the less constraint
ensemble, we are in a situation of ensemble equivalence. The study
of the unconstrained variational problems is then sufficient \cite{BB:2005_JSP}.%
\begin{comment}
: all solutions of the constrained variational problem are solutions
of the unconstrained variational problem. 
\end{comment}
{} We will thus consider, by relaxing one or both constraints, two dual
problems : a) canonical, by relaxing the energy constraint, with the
free energy $F(\beta,\Gamma)=\min_{q}\left\{ -\mathcal{S}[q]\ +\beta\ \mathcal{E}[q]\ |\ \mathcal{C}[q]=\Gamma\right\} $;
b) grand canonical, by relaxing both energy and circulation constraints,
with the thermodynamical potential $J(\beta,\gamma)=\min_{q}\left\{ -\mathcal{S}[q]\ +\beta\ \mathcal{E}[q]+\gamma\mathcal{C}\left[q\right]\right\} $.
%
\begin{comment}
Let us be more precise in the case of the equivalence between the
canonical and the microcanonical ensemble. $E\left(\beta\right)$
is the energy of the canonical equilibria. Then if the ensemble of
all values of $E\left(\beta\right)$ (the range of $E$ in the canonical
ensemble) is the same as the ensemble of all accessible energy, then
both ensembles are equivalent.
\end{comment}
{}It is natural to consider first the grand canonical ensemble. If ensemble
inequivalence does exist, we will then study a more constrained variational
problem, and so on, until the whole range of $E$ and $\Gamma$ has
been covered. Notice that in our case, the ensemble of accessible
values for $\left(\mathcal{E},\ \mathcal{C}\right)$ is the half plane
$E\geq0$. %
\begin{comment}
As will become clear latter, the microcanonical ensemble will be equivalent
to the canonical one, and it will not be studied explicitly.
\end{comment}
{}

\paragraph{Solutions of quadratic variational problems}

For all variational problem to be considered in the following, we
will look for the minimum of a quadratic functional, with a linear
part. Let us call $Q$ the purely quadratic part and $L$ the linear
part of this functionnal. Then we have three cases 

\begin{enumerate}
\item {\small The smallest eigenvalue of $Q$ is strictly positive. The
minimum exists and is achieved by a unique minimizer.}{\small \par}
\item {\small At least one eigenvalue of $Q$ is strictly negative. There
is no minimum (the infimum is $-\infty$).}{\small \par}
\item {\small The smallest eigenvalue of $Q$ is zero (with eigenfunction
$e_{0}$). If $Le_{0}=0$ (case 3a), the maximum exists and each state
of the neutral direction $\left\{ \alpha e_{0}\right\} $ is a minimizer.
If $Le_{0}\neq0$, (case 3b) then no minimum exist. }{\small \par}
\end{enumerate}

\paragraph{The grand canonical ensemble}

We consider in this part $h=0$, so the energy is purely quadratic.%
\begin{comment}
Modification added in presence of topography will be discussed later.
\end{comment}
{} We look for the minimum of $\mathcal{J}=-\mathcal{S}\ +\beta\ \mathcal{E}+\gamma\mathcal{C}$.
%
\begin{comment}
Using the expression of \textbackslash{}mathcal\{S\}, \textbackslash{}mathcal\{E\}
and \textbackslash{}mathcal\{C\},
\end{comment}
{}We introduce the projections $q_{i}$ of the PV $q$ on a complete
orthonormal basis of eigenfunctions $e_{i}(x,y)$ of the operator
$\mathcal{O}$ ($\mathcal{O}[e_{i}]=e_{i}/\lambda_{i}$) (see the
definition of the energy)%
\begin{comment}
with $-\Delta e_{i}=\lambda_{i}e_{i}$
\end{comment}
{} and find:\[
\mathcal{J}[q]=\sum_{i,j\ge1}\delta_{ij}\left(1-\beta/\lambda_{i}\right)q_{i}q_{j}+\sum_{i\ge1}\gamma\langle e_{i}\rangle q_{i}\]

where $\delta_{ij}$ is the Kronecker symbol, and where the $\lambda_{i}$
(all negative) are in decreasing order. One can see that the quadratic
part is diagonal, and that all eigenvalues are strictly positive if
and only if $\beta>\lambda_{1}$(case 1. above) . If $\beta=\lambda_{1}$(case
3), a neutral direction exists if and only if $\gamma=0$ (case 3a).
Thus, grand canonical solutions exist only for $\beta>\lambda_{1}$
or ($\beta=\lambda_{1}$ and $\gamma=0$).%
\begin{comment}
(using the previous paragraph).
\end{comment}
{} By computing the energy and circulation of all those states, we prove
that it exists a unique solution at each point in the diagram $(E,\Gamma)$,
below a parabola $\mathcal{P}$ of equation $E=-\Gamma^{2}\lambda_{1}/\left(2\left\langle e_{1}\right\rangle ^{2}\right)$.
Because the values of energies above the parabola $\mathcal{P}$ are
not reached, we conclude that there is ensemble inequivalence for
parameters in this area. As explained previously, we have to consider
a more constrained variational problem to find solutions in this area.

\paragraph{The canonical ensemble}

We now look for the minimum of $-\mathcal{S}\ +\beta\ \mathcal{E}\ $
for a fixed circulation $\Gamma$. In order to take into account this
constraint, we express one coordinate in term of the others : $q_{1}=\left(\Gamma-\sum_{i}q_{i}\langle e_{i}\rangle\right)/\left\langle e_{1}\right\rangle $.
Then,%
\begin{comment}
using the expression of $\mathcal{E}$ and $S$,
\end{comment}
{} we have to minimize

\begin{eqnarray*}
\mathcal{F}[q] & = & \sum_{i,j\ge2}\left(\delta_{ij}\left(1-\frac{\beta}{\lambda_{i}}\right)+\left(1-\frac{\beta}{\lambda_{1}}\right)\frac{\langle e_{i}\rangle\langle e_{j}\rangle}{\langle e_{1}\rangle^{2}}\right)q_{i}q_{j}\\
 &  & -\sum_{i\ge2}\Gamma\frac{\langle e_{i}\rangle}{\langle e_{1}\rangle{}^{2}}\left(1-\frac{\beta}{\lambda_{1}}\right)q_{i},\end{eqnarray*}
without constraint.

The linear operator $Q$, associated to the quadratic part of $\mathcal{F}$,
is not diagonal in the basis $\left\{ e_{i}\right\} $. We first notice
that if the domain geometry admits one or more symmetries, it generically
exists eigenfunctions having the property $\left\langle e_{i}\right\rangle =0$%
\begin{comment}
(this may also happen in domains without symmetry but it would not
be generic)
\end{comment}
{}. In the subspace spanned by all those eigenfunctions, $Q$ is diagonal,
and its smallest eigenvalue is positive as long as $\beta>\beta_{1}^{0}$,
where $\beta_{1}^{0}$ is the greatest $\lambda_{i}$ on this subspace.
Then we look for the value of $\beta$ such that the smallest eigenvalue
of $Q$ is zero in the subspace spanned by eigenfunctions with $\left\langle e_{i}\right\rangle \neq0$.
Let us call $\beta_{c}^{*}$ this value, and $q_{c}^{*}$ the corresponding
eigenfunction : $Q{[q}_{c}^{*}]=0$.%
\begin{comment}
Its components satisfy : $\forall i\ge2,\ q_{c\ i}^{*}=-\frac{\langle e_{i}\rangle}{\langle e_{1}\rangle^{2}}\frac{1+\frac{\beta}{\lambda_{1}}}{1+\frac{\beta}{\lambda_{i}}}\sum_{j\ge2}\langle e_{j}\rangle q_{c\ j}^{*}$.
We deduce from this expression that $\sum_{j\ge2}\langle e_{j}\rangle q_{c\ j}^{*}\ne0$
(if not, each term $\omega_{c\ i}^{*}$ is zero). Using this fact,
and the previous expression, we find after some manipulation that
$\beta_{c}^{*}$is the greatest zero of the function :
\end{comment}
{} Using this last equation, we prove that $\beta_{c}^{*}$ is the greatest
zero of the function $f(x)=1-x\sum_{i\ge1}\langle e_{i}\rangle^{2}/(x-\lambda_{i})$
.%
\begin{comment}
We show easily that $\max{\beta_{1}^{0},\ \beta_{c}^{*}}<-\lambda_{1}$
\end{comment}
{} %
\begin{comment}
(more negative temperature, and thus higher energy, are accessible
in the canonical ensemble than in the grand canonical ensemble)
\end{comment}
{}We conclude that there is a single solution to the variational problem
for $\beta>\max\left(\beta_{1}^{0},\ \beta_{c}^{*}\right)$ (case
1) and no solution for $\beta<\max\left(\beta_{1}^{0},\ \beta_{c}^{*}\right)$
(case 2). \emph{}When $\beta=\max\left(\beta_{1}^{0},\ \beta_{c}^{*}\right)$,
to discuss the existence of a neutral direction, we distinguish two
cases according to the sign of $\beta_{1}^{0}-\beta_{c}^{*}$ \emph{}: 

\begin{description}
\item [{{\small i)}}] {\small $\beta_{1}^{0}<\beta_{c}^{*}$ we then consider
$\beta=\beta_{c}^{*}$. We are in case 3a for $\Gamma=0$ and in case
3b for $\Gamma\neq0$. }{\small \par}
\item [{{\small ii)}}] {\small $\beta_{1}^{0}>\beta_{c}^{*}$ we then consider
$\beta=\beta_{1}^{0}$. We are in case 3a whatever the value of $\Gamma$. }{\small \par}
\end{description}
In case i), if $\Gamma\ne0$, %
\begin{comment}
because $\lim_{\beta\rightarrow\beta_{c}^{*}}E\left(\beta\right)=+\infty$,
\end{comment}
{}all energy value are reached in the canonical ensemble. If $\Gamma=0$,
the solutions of the neutral directions are $\alpha q_{c}^{*}$. Thus,
varying $\alpha$, all energy values are reached ; there is two canonical
solutions corresponding to each energy $E$, depending on the sign
of $\alpha$. In case ii), $\lim_{\beta\rightarrow\beta_{1}^{0}}E(\beta)=E_{\mathcal{P}_{0}}(\Gamma)\propto\Gamma^{2}$%
\begin{comment}
$\lim_{\beta\rightarrow\beta_{1}^{0}}E(\beta)=E_{\mathcal{P}_{0}}(\Gamma)=\Gamma^{2}\left(\sum_{i\ge1}\frac{\langle e_{i}\rangle^{2}}{\lambda_{i}}\left(\frac{1}{\beta_{1}^{0}}+\frac{1}{\lambda_{i}}\right)^{-2}\right)/\left(\sum_{i\ge1}\langle e_{i}\rangle^{2}\left(\frac{1}{\beta_{1}^{0}}+\frac{1}{\lambda_{i}}\right)^{-2}\right)$ 
\end{comment}
{}. This defines a parabola $\mathcal{P}_{0}$%
\begin{comment}
in the plane $\left(E,\Gamma\right)$
\end{comment}
{}%
\begin{comment}
(a direct computation gives the the curvature of the parabola which
on the domain geometry)
\end{comment}
{}. Then, whatever the value of $\Gamma$, there is a unique canonical
solution $q\left(\beta,\Gamma\right)$ for each point of the diagram
$(E,\Gamma)$ below $\mathcal{P}_{0}$. The canonical solutions of
the neutral directions is $q=\alpha e_{1}^{0}+\lim_{\beta\rightarrow\beta_{1}^{0}}q\left(\beta,\Gamma\right)$
where $e_{1}^{0}$ is the eigenfunction associated with $\beta_{1}^{0}$.
Varying $\alpha$, all energy larger than $E_{\mathcal{P}_{0}}(\Gamma)$
are reached. For each energy above the parabola $\mathcal{P}_{0}$,
it exists two canonical solutions, depending on the sign of $\alpha$.
In both cases we find that all circulation and energy values have
been reached by canonical solutions. We conclude that microcanonical
and canonical ensembles are equivalent. %
\begin{comment}
: all microcanonical solutions are also canonical solutions.
\end{comment}
{}

\paragraph{Description \emph{}of \emph{}phase \emph{}transitions\emph{. }}

\begin{figure}
\resizebox{8truecm}{!}{\includegraphics{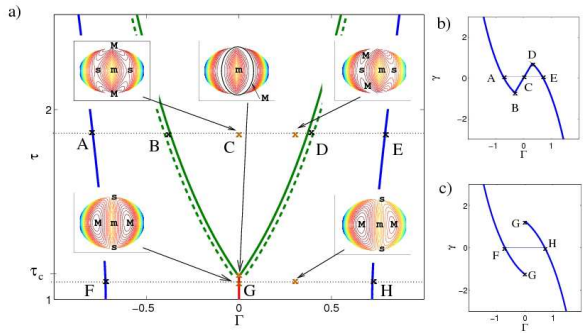}}

\caption{First observation of a bicritical point (change from a first order
to two second order phase transitions) in a system with long range
interactions. Euler equation in a rectangular domain of aspect ratio
$\tau$. Green line : second order phase transition (discontinuity
of $\partial\gamma/\partial\Gamma$ in b). Red line : 1st order phase
transition (discontinuity of $\gamma$ in c). Insets are projection
of the Entropy $\mathcal{S}[q]$ in a plane $(q_{1}^{0},q_{1})$ for
fixed energy and circulation. }

\label{figplotDdep} 
\end{figure}

In case i), at fixed energy, we can show by a direct computation that
$\gamma(\Gamma)=\frac{\partial S}{\partial\Gamma}$ %
\begin{comment}
is a decreasing function if $\Gamma\ne0$, and
\end{comment}
{} is discontinuous in $\Gamma=0$. This means that there is a microcanonical
first order transition (see point G figure \ref{figplotDdep}-c).
In case ii), one can show %
\begin{comment}
that $\gamma(\Gamma)$ is continuous everywhere and 
\end{comment}
{} that $\frac{\partial\gamma}{\partial\Gamma}$ is discontinuous when
$(E,\Gamma)$ belongs to $\mathcal{P}_{0}$ (see points B and D, figure
\ref{figplotDdep}-b). \emph{}The ensemble inequivalence area is associated
with the existence of a first order transition in the ensemble with
only one constraint on the energy (see the corresponding Maxwell constructions
on figure \emph{}\ref{figplotDdep}-b,c). Similarly, if we now fix
the circulation, there is a discontinuity of $\frac{\partial\beta}{\partial E}$
when $(E,\Gamma)$ belongs to $\mathcal{P}_{0}$. It means that $\mathcal{P}_{0}$
is a line of second order phase transition for canonical and microcanonical
ensembles. %
\begin{comment}
In the semi plane $(E,\Gamma)$, the choice of one state among the
two possibilities above the parabola $\mathcal{P}_{0}$ breaks the
symmetry (this is shown figure \ref{fig:Phase_Transition_General_domain},
in presence of topography)
\end{comment}
{}

\paragraph{General criteria.}

The main interest of the abstract previous analysis is to conclude
that all of the models considered will behave according to only two
types of phase diagram structures. The difference between the two
classes of systems is the existence of either first or second order
phase transitions, corresponding respectively to case i) and ii).
%
\begin{comment}
It is thus essential to analyze the criteria for class i) or class
ii) systems.
\end{comment}
{}If the domain has no symmetry, only case i) is possible (generically).
If there is a symmetry, both cases are possible. The criteria for
class i) or class ii) systems is the sign of $\beta_{1}^{0}-\beta_{c}^{*}$,
that can be easily computed for a system at hand. The same criteria
was obtained in \cite{ChavanisSommeria:1996_JFM_Classification} for
the Euler equation, using a different method. Very interestingly,
this criteria does not depend on the model considered (it does not
depend on the Rossby radius of deformation $R$), but only on the
domain shape. \textbf{}For some classes of domain geometries, as ellipses
for instance, we are always in case ii). More generally, any domain
geometry sufficiently stretched in a direction perpendicular to its
symmetry axis is in case ii). This is for instance the case of a rectangular
domain The transition from systems of type i) to systems of type ii),
when the geometry is modified, leads to very interesting phenomena
that are described now.%

\paragraph{The bicritical point}

On figure \ref{figplotDdep}, we consider a fixed energy, and present
the phase diagram in the $\left(\Gamma,\tau\right)$ plane, where
$\tau$ is a parameter characterizing the aspect ratio of the domain
(horizontal over vertical width). In the microcanonical ensemble,
there is a bifurcation from a first order transition line to two second
order transition lines at a critical value $\tau=\tau_{c}$. Such
a bifurcation is referred as a bicritical point (see \cite{BB:2005_JSP}).
%
\begin{comment}
As explained in \cite{BB:2005_JSP}, many other generic phase transitions
have never been observed. 
\end{comment}
{}

\paragraph{With a topography}

We now give the main striking features arising when adding the term
$h$, especially the new unobserved phase transitions. %
\begin{comment}
Because of this new term, the operator $L$ associated with the linear
part of the functional to minimize is changed , but the operator $Q$
associated with its quadratic part is left unchanged.
\end{comment}
{} Concerning the existence of first and second order phase transitions
in the microcanonical ensemble, there is now three possibilities.
If $\beta_{1}^{0}-\beta_{c}^{*}<0$, the phase diagram is similar
to the one of case i). If $\beta_{1}^{0}-\beta_{c}^{*}>0$ and $\left\langle h,e_{1}^{0}\right\rangle =0$,
then we are in the case ii), except that the minima of $\mathcal{P}$
and $\mathcal{P}_{0}$ are no more at the same place in the diagram
$(E,\Gamma)$. If $\beta_{1}^{0}-\beta_{c}^{*}>0$ and $\left\langle h,e_{1}^{0}\right\rangle \ne0$,
there is neither second nor first order transition. %
\begin{comment}
Whatever the topography, the three cases are possible, depending on
the domain geometry. The solution at high energy is a monopole proportional
to $q_{c}^{*}$ in case i), and a dipole proportional to $e_{1}^{0}$
in case ii) and iii).
\end{comment}
{}%
\begin{comment}
As an example, let us consider an arbitrary domain geometry, with
an horizontal (x direction) and a vertical symmetry, that fall into
case i), choose $h=-\beta_{cor}y$. At low energy, the flow is the
usual Fofonoff mode, with the stream function proportional to the
topography in the inner domain, and strong recirculating jets at the
boundaries. If the domain is sufficiently stretched in the x direction,
there is a phase transition where a dipole in the x direction is superposed
to the initial dipole in the y direction . If the domain is sufficiently
stretched in the y direction, the solution at high energy is still
a dipole having the Fofonoff mode structure, but with no strong jets
confined at the boundary. 
\end{comment}
{}

\paragraph{Second order Azeotropy}

\begin{figure}
\resizebox{8truecm}{!}{\includegraphics{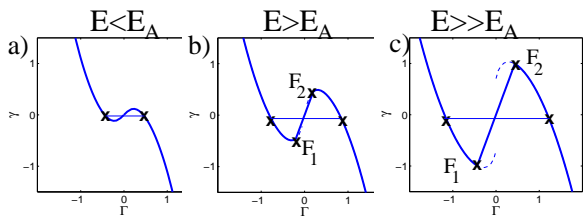}}

\caption{First observation of second order azeotropy \textbf{(}{\small at
energy $E_{A}$ there is simultaneous appearance of two second order
phase transitions). a) $\partial\gamma/\partial\Gamma$ is continuous
; b and c) $\partial\gamma/\partial\Gamma$ is not continuous in $F_{1}$
and $F_{2}$. }}

\label{fig:azeotropy} 
\end{figure}

In case ii) with topography, if we consider the energy as an external
parameter, there is the simultaneous appearance of two second order
phase transitions in the microcanonical ensemble (see figure \ref{fig:azeotropy}
and the corresponding flows figure \ref{fig:Phase_Transition_General_domain}),
which is the signature of second order azeotropy. %
\begin{comment}
The possible existence of such a phase transition, called second order
azeotropy, was predicted by \cite{BB:2005_JSP} but never observed
in any physical system.
\end{comment}
{} %
\begin{comment}
This is to our knowledge the first example of azeotropy in a system
with long range interactions.
\end{comment}
{}

\paragraph{Conclusion}

We report the generic existence of ensemble inequivalence and of new
phase transitions in a large class of 2D flows. All phase transitions
presented here appear in the inequivalence ensemble area. They are
in that respect a signature of such an inequivalence in a long range
interacting system, which has never been observed experimentally.
The observation of those transitions could be carried in laboratory
experiments on quasi 2D flows. This could be done by using either
magnetized electron columns \cite{Schecter_Dubin_etc_Vortex_Crystals_2DEuler1999PhFl}
or three dimensionnal tanks with small height compared to the horizontal
scale, with a further ordering (strong rotation or a transverse magnetic
field). %
\begin{comment}
The results are presented here for one layer models, but it is possible
to use the same method to study multi layers QG models. Strikingly,
this lead to the same criteria on the domain geometry for the existence
of a second order phase transition.
\end{comment}
{}The interest for ocean applications will be described in a companion
paper, as well as the detailed complete computations, generalization
to multi level QG equations and analysis of the ensemble inequivalence
associated to the bicritical points. %
\begin{comment}
In the latter case, effects of small dissipation and forcing should
be taken into account. We will discuss this in detailed in future
works.
\end{comment}
{} 

\acknowledgments We thank J. Barré, P.H. Chavanis and J. Sommeria
for interesting discussions. This work was supported by the ANR program
STATFLOW ( ANR-06-JCJC-0037-01 ).

\bibliography{FBouchet,Long_Range,Meca_Stat_Euler,Ocean,Experimental_2D_Flows}

\end{document}